\documentclass[12pt]{article}

\newcommand{\be}{\begin{equation}}
\newcommand{\ee}{\end{equation}}
\newcommand{\bel}[1]{\begin{equation}\label{#1}}
\newcommand{\bea}{\begin{eqnarray}}
\newcommand{\eea}{\end{eqnarray}}
\newcommand{\ba}{\begin{array}}
\newcommand{\ea}{\end{array}}

\newcommand{\nn}{\nonumber \\}
\newcommand{\bra}[1]{\mbox{$\langle \, {#1}\, |$}}
\newcommand{\ket}[1]{\mbox{$| \, {#1}\, \rangle$}}
\newcommand{\exval}[1]{\mbox{$\langle \, {#1}\, \rangle$}}

\def\bbbc{{\mathchoice {\setbox0=\hbox{$\displaystyle\rm C$}\hbox{\hbox
to0pt{\kern0.4\wd0\vrule height0.9\ht0\hss}\box0}}
{\setbox0=\hbox{$\textstyle\rm C$}\hbox{\hbox
to0pt{\kern0.4\wd0\vrule height0.9\ht0\hss}\box0}}
{\setbox0=\hbox{$\scriptstyle\rm C$}\hbox{\hbox
to0pt{\kern0.4\wd0\vrule height0.9\ht0\hss}\box0}}
{\setbox0=\hbox{$\scriptscriptstyle\rm C$}\hbox{\hbox
to0pt{\kern0.4\wd0\vrule height0.9\ht0\hss}\box0}}}}

\def\bbbq{{\mathchoice {\setbox0=\hbox{$\displaystyle\rm
Q$}\hbox{\raise 0.15\ht0\hbox to0pt{\kern0.4\wd0\vrule 
height0.8\ht0\hss}\box0}}
{\setbox0=\hbox{$\textstyle\rm Q$}\hbox{\raise
0.15\ht0\hbox to0pt{\kern0.4\wd0\vrule height0.8\ht0\hss}\box0}}
{\setbox0=\hbox{$\scriptstyle\rm Q$}\hbox{\raise
0.15\ht0\hbox to0pt{\kern0.4\wd0\vrule height0.7\ht0\hss}\box0}}
{\setbox0=\hbox{$\scriptscriptstyle\rm Q$}\hbox{\raise
0.15\ht0\hbox to0pt{\kern0.4\wd0\vrule height0.7\ht0\hss}\box0}}}}

\def\bbbt{{\mathchoice {\setbox0=\hbox{$\displaystyle\rm
T$}\hbox{\hbox to0pt{\kern0.3\wd0\vrule height0.9\ht0\hss}\box0}}
{\setbox0=\hbox{$\textstyle\rm T$}\hbox{\hbox
to0pt{\kern0.3\wd0\vrule height0.9\ht0\hss}\box0}}
{\setbox0=\hbox{$\scriptstyle\rm T$}\hbox{\hbox
to0pt{\kern0.3\wd0\vrule height0.9\ht0\hss}\box0}}
{\setbox0=\hbox{$\scriptscriptstyle\rm T$}\hbox{\hbox
to0pt{\kern0.3\wd0\vrule height0.9\ht0\hss}\box0}}}}

\def\bbbs{{\mathchoice
{\setbox0=\hbox{$\displaystyle     \rm S$}\hbox{\raise0.5\ht0\hbox
to0pt{\kern0.35\wd0\vrule height0.45\ht0\hss}\hbox
to0pt{\kern0.55\wd0\vrule height0.5\ht0\hss}\box0}}
{\setbox0=\hbox{$\textstyle        \rm S$}\hbox{\raise0.5\ht0\hbox
to0pt{\kern0.35\wd0\vrule height0.45\ht0\hss}\hbox
to0pt{\kern0.55\wd0\vrule height0.5\ht0\hss}\box0}}
{\setbox0=\hbox{$\scriptstyle      \rm S$}\hbox{\raise0.5\ht0\hbox
to0pt{\kern0.35\wd0\vrule height0.45\ht0\hss}\raise0.05\ht0\hbox
to0pt{\kern0.5\wd0\vrule height0.45\ht0\hss}\box0}}
{\setbox0=\hbox{$\scriptscriptstyle\rm S$}\hbox{\raise0.5\ht0\hbox
to0pt{\kern0.4\wd0\vrule height0.45\ht0\hss}\raise0.05\ht0\hbox
to0pt{\kern0.55\wd0\vrule height0.45\ht0\hss}\box0}}}}

\def\bbbz{{\mathchoice {\hbox{$\sf\textstyle Z\kern-0.4em Z$}}
{\hbox{$\sf\textstyle Z\kern-0.4em Z$}}
{\hbox{$\sf\scriptstyle Z\kern-0.3em Z$}}
{\hbox{$\sf\scriptscriptstyle Z\kern-0.2em Z$}}}}

\def\gsim {\mbox{\hbox{ \lower-.6ex\hbox{$>$}
\kern-1.12em \lower.5ex\hbox{$\sim$}\kern+.35em}}}
\def\lsim {\mbox{\hbox{ \lower-.6ex\hbox{$<$}
\kern-1.12em \lower.5ex\hbox{$\sim$}\kern+.35em}}}

\begin{document}
\begin{center}          
{\large                       
{\bf Shocks in the asymmetric exclusion process with internal
degree of freedom}
}\\[3cm]
{\large {
Fatemeh Tabatabaei\footnote{f.tabatabaei@fz-juelich.de}$^{\dag}$ and Gunter M. Sch\"utz\footnote{g.schuetz@fz-juelich.de}$^{\dag\ddag}$
}}\\[8mm]
{\em \dag Institut f\"ur Festk\"orperforschung, Forschungszentrum J\"ulich,\\
52425 J\"ulich, Germany}\\[2mm]
{\em \ddag Interdisziplin\"ares Zentrum f\"ur komplexe Systeme \\
University of Bonn, Germany}\\[4mm]
    
\vspace{1cm}         
\begin{minipage}{13cm}{
\baselineskip 0.3in
We determine all families of Markovian three-states lattice gases with pair 
interaction and a single local conservation law. One such family of models 
is an asymmetric exclusion process where particles
exist in two different nonconserved states. We derive conditions 
on the transition rates between the two states such that the shock
has a particularly simple structure with minimal intrinsic
shock width and random walk dynamics. We calculate the drift 
velocity and diffusion coefficient of the shock.\\[4mm]
}
\end{minipage} \end{center} 


\newpage
\baselineskip 0.3in
\section{Introduction}
\label{intro}

The asymmetric simple exclusion process (ASEP) \cite{Ligg99,Schu00}
has often been called the Ising model of nonequilibrium statistical
physics. In this stochastic lattice gas model particles move randomly
with a bias onto neighboring lattice sites, provided the target
site is empty. Even its most-studied one-dimensional version 
which describes driven single-file diffusion
exhibits rich phenomena, in particular boundary-induced phase 
transitions \cite{Krug91,Schu93,Derr93,Kolo98,Popk99}, and has a wide range of 
applications in different branches of physics. Experiments probing 
single-file diffusion have been performed with molecules in zeolites 
\cite{Kukl96}, colloidal particles in confined geometry \cite{Wei00} 
or optical lattices \cite{Lutz04}, and very recently with granular 
systems \cite{Coup06}. Driven single-file transport has been observed in 
biological systems and the ASEP serves as a starting point for modelling 
the motion of ribosomes along the m-RNA during protein synthesis 
\cite{MacD68,Schu97} and of molecular motors along microtubuli or 
actin filaments \cite{Nish05}. The ASEP is also at the heart of the
cellular automaton approach to vehicular traffic on road networks
\cite{Chow00,Helb01}. In this setting some predictions from the
theory of boundary-induced phase transitions
for the development of traffic jams have been verified empirically 
despite the complexity of real vehicular traffic \cite{Popk01}. 
Phenomena analogous to traffic jams have also been observed in the 
biological transport systems referred to above. 

In the hydrodynamic approach to traffic flow \cite{Helb01} using partial 
differential equations traffic jams correspond to shock solutions. 
A shock is a density discontinuity on moving with some deterministic speed,
determined by mass conservation. It is therefore no surprise that
on macroscopic Euler scale the time evolution of the
particle density of the ASEP is described by the inviscid
Burgers equation \cite{Burg74,Kipn99} which develops shocks
for generic initial data. With a view on applications of the ASEP
to systems for which a hydrodynamic description is too coarse-grained it
would thus be of interest to understand what fluctuating microscopic 
structure (on lattice scale) is underlying the phenomenon of shocks.

In fact, a great deal is known about shocks in the ASEP due to the exact 
solubility of the model. In the stationary regime the shock structure has been 
studied as seen from a so-called second-class particle which serves as
microscopic marker of the shock position. The particle density decays away 
from the shock exponentially (on lattice scale) to the respective constant 
bulk values $\rho_{1,2}$ of the two branches of the shock 
\cite{Ferr91,Derr97,Derr98}. The shock position itself has been proved
to perform Brownian motion on coarse grained diffusive scale \cite{Ferr94}.
For a particular strength of the driving
field the associated decay constant of the particle density
vanishes, corresponding to a ``minimal'' intrinsic shock width. 
For this special value of the driving field also the motion
of the shock simplifies greatly. It performs a biased random
walk on microscopic {\it lattice} scale with explicitly known 
hopping rates \cite{Beli02,Kreb03}.

It is natural to ask whether this special feature of the ASEP
survives in more complicated models of driven diffusive systems.
In particular, one would like to investigate exclusion
processes with nonconserved internal degrees of freedom, where 
particles may have different velocities, charges, masses or
other distinguishing properties that they can gain or lose 
e.g. in a collision or chemical reaction. Here we address this
question in the simplest case of two possible internal states
that each particle can posses. Such models have been investigated
recently for various biological and vehicular transport phenomena
\cite{Nish05,Oloa98,Nish03}. Studying the microstructure of a
shock illuminates the role of finite-size effects in first-order
boundary-induced phase transitions that are associated with
the motion of traffic jams \cite{Dudz00,Nagy02,Gier05} in
finite systems.

The paper is organized as follows: In the following section we determine 
the families of three-states models with pair interaction and
a single conservation law. We also define shock measures for these systems. 
In Sec.~3 we study exclusion processes with binary internal degree of freedom
that allows for special travelling shock solutions on the finite lattice. 
In Sec.~4 we summarize our results and draw some conclusions.

\section{Three-states processes with one conservation law}
\label{Sec2}

\subsection{Stochastic dynamics}

On an abstract level the exclusion process with a binary internal degree
of freedom is a three-states process where the state of the system at 
any given time is described by a set of ``occupation 
numbers'' $\underline{n}={n_1,\dots,n_L}$ where $n_k=0,1,2$ is the local 
occupation number
at site $k$ and $L$ is the number of sites. In the next section we assign 
state 0 to an empty lattice site, state 1 to a particle of type $A$ and 
state 2 to a particle of type $B$. The labels $A$ and $B$ represent two 
possible internal
states of a particle. However, in this section we first keep the
treatment general and consider the occupation numbers as abstract objects,
labelling one out of three possible states of a lattice site.

The bulk stochastic dynamics
are defined by nearest neighbor transitions between the occupation variables
which occur independently and randomly in continuous time after an 
exponentially distributed waiting time. The mean 
$\tau(n_k',n_{k+1}';n_k,n_{k+1})$ of this waiting time depends on the 
transition $(n_k,n_{k+1}) \to (n_k',n_{k+1}')$, its inverse is the
transition rate.

At the boundary sites $1,L$ we assume the system to be connected to
some external reservoir with which the system can exchange
particles. 
For injection and extraction of particles at the left boundary we introduce 
the
rates :
\begin{eqnarray}
  0 \rightleftharpoons 1
   \quad & \alpha_1, \; \gamma_1, \nonumber \\
  0\rightleftharpoons 2 
  \quad & \alpha_2,\; \gamma_2,  \nonumber \\   
 1 \rightleftharpoons 2
   \quad & \alpha_3, \; \gamma_3,    
\end{eqnarray}
and for the right boundary 
\begin{eqnarray}
  0 \rightleftharpoons 1
   \quad & \delta_1, \; \beta_1, \nonumber \\
  0\rightleftharpoons 2 
  \quad & \delta_2,\; \beta_2,  \nonumber \\   
 1 \rightleftharpoons 2
   \quad & \delta_3, \; \beta_3.   
\end{eqnarray}
Here and below the left rate refers to the process going from
left to right, while the right rate is for the reversed process.

The time evolution defined above can be written in terms of a continuous-time 
master equation for the
probability vector
\begin{equation}
\ket{P(t)} = \sum_{\underline{n}} P(n_1, \cdots, n_L;t)\ket{\underline{n}},
\end{equation}
where $P(n_1, \cdots, n_L;t)$ is the distribution for the probability of 
finding particles at sites 1 to $L$ and $\ket{\underline{n}}$ is the basis 
vector in 
the space of configurations in the naturally defined
tensor basis \cite{Schu00}. The probability vector is 
normalized such that 
$\langle s | P \rangle =1$ with the summation vector 
$\bra{s}=\sum_{\underline{n}}\bra{\underline{n}} $ and scalar product
$\langle \underline{n} | \underline{n}' \rangle =
\delta_{\underline{n},\underline{n}'}$. 
The time evolution is generated by the stochastic ``quantum Hamiltonian''
$H$ whose offdiagonal matrix elements $H_{\underline{n},\underline{n}'}$ are 
the negative transition rates between configurations. As required by 
conservation of probability, the diagonal elements 
are the negative sum of transition rates in the respective column.

Therefore the master equation is now  
described by the imaginary time Schr\"odinger equation
\begin{equation}
\frac{d}{dt}\ket{P(t)}= -H \ket{P(t)}
\end{equation}
with the formal solution
\bel{2-12}
\ket{P(t)} = \mbox{e}^{-Ht}\ket{P(0)}.
\ee
Since only nearest-neighbour interactions are included, 
the quantum Hamiltonian $H$ defined above has the structure
\be      
H = b_1+\sum_{k=1}^{L-1} h_{k,k+1} + b_L.\label{eq:r30}
\ee
where $b_1$ and $b_L$ are the boundary matrices
\be
b_1 = - \left( \ba{cccc}
 -(\alpha_1+\alpha_2) & \gamma_1 & \gamma_2  \\
\alpha_1 & -(\gamma_1+\alpha_3) & \gamma_3  \\
\alpha_2 & \alpha_3 & -(\gamma_2+\gamma_3) \ea
\right)_1,\label{eq:r28}
\ee 
\vspace{1cm}
\be
b_L = - \left( \ba{cccc}
 -(\delta_1+\delta_2) & \beta_1 & \beta_2  \\
\delta_1 & -(\beta_1+\delta_3) & \beta_3  \\
\delta_2 & \beta_3 & -(\beta_2+\beta_3) \ea
\right)_L\label{eq:r29}.
\ee

The local bulk transition matrix $h_{k,k+1}$ acts non-trivially only on sites 
$k$ and $k+1$. To define its matrix elements we introduce an integer label 
\bel{label}
i=3n_k+n_{k+1}+1
\ee
in the range $1 \leq i \leq 9$ for the occupation variables on two 
neighboring sites. The offdiagonal matrix elements $(h_{k,k+1})^{(ij)}$ 
are then the transition rates $-w_{ij}$. Here $i=3n_k'+n_{k+1}'+1$ 
labels the target configuration and $j$ is the respective label 
of the initial configuration $(n_k,n_{k+1})$.

\subsection{Symmetries and conservation laws}

Within this setting one could describe 72 different bulk transitions, 
corresponding to the 72 mathematically possible changes of configurations
on a pair of sites. However, we shall reduce this large number by imposing 
various physically motivated constraints. First, we require a local 
conservation law. Generally, the physical 
interpretation of the conservation law depends on the physical 
interpretation of the occupation numbers $n_k$ and will become clear below.
Mathematically this means that in a periodic system
some function $\sum_k C(n_k)$ of the local occupation numbers 
should remain invariant
under the stochastic dynamics, i.e.,
\bel{conservation}
C(n_k')+C(n_{k+1}') = C(n_k)+ C(n_{k+1})
\ee
for any local transition between configurations $i,j$. This constraint
forces a large number of transition
rates $w_{ij}$ to vanish. Physically $C(n)$ is some
observable property (such as mass or charge) of the state $n$.

The conservation condition (\ref{conservation}) does not
uniquely define the function $C(n)$.
In order to analyse these constraints we set $C(0)=0$ and 
$C(1)=1$. This involves no loss of generality since adding
a constant to $C(n)$ or multiplying $C(n)$ by an arbitrary factor leaves
(\ref{conservation}) invariant.
From inspection of (\ref{conservation}) one can then see that there are three
distinct families of solutions: (i) degenerate case, represented by 
$C(2)=C(1)=1$ 
(or equivalently $C(2)=C(0)=0$), (ii) linear nondegenerate case, represented by 
$C(2)=2$ 
(or equivalently $C(2)=-1$, $C(2)=1/2$), (iii) two independent conservation
laws, represented by any other value of $C(2)$. The non-degenerate linear
conservation law is treated elsewhere \cite{Taba06b}, the case of 
two conservation laws was studied in detail in \cite{Rako04,Jafa05}. Here we 
investigate the
degenerate conservation law. The degenerate function $C(n)$ has a natural
interpretation as counting the number of particles at a given site
irrespective of its internal state. This is the motivation behind the
assignment of the state labels $A,B$ used below.

The presence of a conservation law implies a lattice continuity equation
\bel{continuity}
\frac{d}{dt} C_k = j_{k-1} - j_k                              
\ee
for the expectation $C_k = \exval{C(n_k)}$. This quantity plays the role
of a local order parameter. The quantity $j_k$ is the current
associated with the conservation law. It is given by the expectation
of some combination of local occupation numbers, depending on the model under
investigation, see below. Since we do not study here periodic systems
we do not require the boundary sites where the system is connected to
the reservoir to respect the conservation law. The quantities $j_0$, $j_L$
entering the continuity equation for $k=1$ and $k=L$ respectively
are source terms resulting from the 
reservoirs. They are functions of the reservoir densities.

Second, in addition to the conservation law we require $PT$-invariance, 
i.e., the bulk dynamics should be symmetric under 
combined time reversal and space reflection. This physical input generalizes 
the equilibrium condition of detailed balance to allow for external driving 
forces 
which lead to a bias in the hopping rates. In such a case the system is forced 
into a nonequilibrium steady state with a stationary current flowing in the 
system. Well-known examples for models of this kind are exclusion processes 
satisfying pairwise balance \cite{Schu96}. $PT$-invariance is implemented by
demanding detailed balance with respect to the reflected target state of a
local transition. As a result, there are pairwise relations 
between some of the 72 transition rates, see below.

\subsection{Product measures}

Even though the number of independent model parameters is greatly reduced 
by particle conservation and $PT$-symmetry the form of the
stationary distribution is not determined by these constraints.
In order to able to carry out explicit computations we 
restrict ourselves to systems such that the stationary
distribution of the stochastic dynamics factorizes, i.e., one has
a product measure.
In the quantum Hamiltonian formalism introduced above a
product measure is given by a tensor product
\begin{equation}
\ket{P}=|P_1)\otimes |P_2)\otimes ... \otimes |P_L).
\end{equation}
Here the three-component single-site probability vectors $|P_k)$ has
as components the probabilities $P(n_k)$ of finding state $n$ at site
$k$. In the stationary distribution these probabilities are
position-independent, $|P_k) \equiv |P)$, and the stationary
probability vector thus has the homogeneous product form
\begin{equation}
\ket{P^*}=|P)^{\otimes L}.
\end{equation}

We represent the single-site basis vectors for this model as
\begin{equation}
|0)=\left( \ba{c} 1 \\ 0\\0 \ea \right),\hspace{5mm}
|A)=\left( \ba{c} 0 \\ 1 \\ 0 \ea \right),\hspace{5mm}
|B)=\left( \ba{c} 0 \\ 0 \\ 1 \ea \right)
\end{equation}
and parametrize the stationary one-site marginal
\begin{equation}\label{marginal}
|P)=\frac{1}{\nu}\left( \ba{c} 1 \\ z\\cz \ea \right)
\end{equation}
by a fugacity $z$ and the ratio $c$
of $A$ and $B$ concentrations. The normalization factor
\bel{partitionfunction}
\nu = 1+z+cz
\ee
is the local partition function. Thus one has for this grandcanonical ensemble
\bel{densities}
\rho^A = \frac{z}{\nu}, \quad \rho^B = c\frac{z}{\nu}
\ee
and for the total conserved particle density
\bel{density}
\rho:=\rho^A+\rho^B= z\frac{d}{dz}\ln{\nu} = (1+c)\frac{z}{\nu}.
\ee
In formal analogy to systems in thermal equilibrium we shall
refer to the logarithm of the fugacity as chemical potential.

By definition of stationarity the stationary probability vector satisfies 
the eigenvalue equation
\begin{equation}
H\ket{P^*}=0 \label{eq:r1}.
\end{equation}
Requiring the existence of a stationary product measure imposes constraints 
both  on the bulk rates and on the boundary rates which fix the bulk
fugacity $z$. Once these conditions are 
determined the model is fully defined and its stationary distribution is given.
Notice that by definition a stationary product measure has no correlations
between the occupation numbers at different sites.

After defining the model in this way we shall relax some of the constraints
on the boundary conditions and study the time evolution of (nonstationary) shock
measures of the form
\begin{equation}
\ket{k}=|P_1)^{\otimes k} \otimes |P_2)^{\otimes L-k}.
\end{equation}
These shock measures have single-site probabilities given by a fugacity
$z_1$ in 
the left chain segment up to site $k$
and fugacity $z_2$ in the remaining
chain segment from site $L-k$ up to site $L$. Such a shock measure
defines fully the internal structure of the shock. Since there 
are no correlations in a shock measure one may regard the lattice unit 
as the intrinsic 
shock width. A typical configuration has a sharp decrease of the mean 
interparticle
distance across the lattice point $k$. The boundary
fugacities of the system are chosen such that each chain segment is stationary
at its boundary. The measure itself, however, is not stationary
for $z_1\neq z_2$. The associated gradient of the chemical
potential together with external driving forces entering
the bulk hopping rates drive the system into an nonequilibrium
steady state, to be determined below as the final stage of the
time evolution of the shock measure.

\section{Exclusion process with binary internal degree of freedom}

We now implement the constraints discussed above. The degenerate conservation law
(\ref{conservation}) forces 48 transition rates to vanish. The following 24 transitions
remain:
\begin{eqnarray}
0A\rightarrow \hspace{2mm} \ A0  \quad & w_{42}, \quad A0\rightarrow \ 0A  
\quad & w_{24},\nonumber\\
0B\rightarrow \hspace{2mm} \ B0  \quad & w_{73}, \quad B0\rightarrow \ 0B  
\quad & w_{37},\nonumber\\
AB\rightarrow \ BA               \quad & w_{86}, \quad BA\rightarrow \ AB  
\quad & w_{68}, \nonumber \\
B0\rightarrow \hspace{2mm} \ A0  \quad & w_{47},\quad  0A\rightarrow \ 0B  
\quad & w_{32},  \nonumber \\ 
0B\rightarrow \hspace{2mm} \ A0  \quad & w_{43}, \quad 0A \rightarrow\ B0 
\quad &  w_{72}, \nonumber  \\
A0\rightarrow \hspace{2mm} \ B0  \quad & w_{74}, \quad 0B \rightarrow\ 0A 
\quad & w_{23}, \nonumber  \\
B0\rightarrow \hspace{2mm} \ 0A  \quad & w_{27}, \quad A0 \rightarrow\ 0B 
\quad & w_{34}, \nonumber\\
BA\rightarrow  \ AA  \quad & w_{58}, \quad  AA \rightarrow\ AB \quad & w_{65}, 
\nonumber\\
AB\rightarrow  \ AA  \quad & w_{56}, \quad  AA \rightarrow\ BA \quad & w_{85}, 
\nonumber\\
BB\rightarrow  \ AA  \quad & w_{59}, \quad  AA \rightarrow\ BB \quad & w_{95}, 
\nonumber\\
BB\rightarrow  \ BA  \quad & w_{89}, \quad  AB \rightarrow\ BB \quad & w_{96}, 
\nonumber\\
BB\rightarrow \ AB  \quad & w_{69}, \quad  BA \rightarrow\ BB \quad & w_{98}. 
\nonumber \label{eq:r27}\\ 
\end{eqnarray}

Parity-time invariance leads to pairwise relations between some of
these rates. Time reversal 
symmetry 
means to have detailed balance 
$p^*(\underline{n})w(\underline{n}\rightarrow \underline{n}')=
p^*(\underline{n}')w(\underline{n}'\rightarrow\underline{n})$. 
In order to combine this relation with
the parity (space reflection) operation we change the 
position of neighbouring sites with each other in the initial configuration 
and  final configuration on the left-hand side of the detailed-balance 
relation. 
Using (\ref{marginal}) this yields the following symbolic relations for the rates
\begin{equation}
w(A\rightarrow\ B )= c w(B\rightarrow A)
\end{equation}
for each particle on a pair of neighboring sites. With this relation
we can reduce the number of independent rates in the 
process (\ref{eq:r27}) to only 15 nonstationary rates, viz. 6 hopping rates and 9 ``reaction
rates'' for changes of the internal states of the particles.
For clarity we represent all of the hopping rates by $h$'s and reaction 
process by $r$'s and write the rates as  
\begin{eqnarray}
 w_{47}=r_1 ,\hspace{1mm} w_{32}=cr_1,\nonumber \\
 w_{43}=r_2 ,\hspace{1mm} w_{72}=cr_2, \nonumber \\
 w_{23}=r_3 ,\hspace{1mm} w_{74}=cr_3, \nonumber \\
 w_{27}=r_4 ,\hspace{1mm} w_{34}=cr_4, \nonumber\\
 w_{58}=r_5 ,\hspace{1mm} w_{65}=cr_5, \nonumber\\
 w_{56}=r_6 ,\hspace{1mm} w_{85}=cr_6,  \nonumber\\
 w_{59}=r_7 ,w_{95}=c^2r_7,\nonumber\\
 w_{89}=r_8 ,\hspace{1mm}w_{96}=cr_8,\nonumber\\
 w_{69}=r_9 ,\hspace{1mm}w_{98}=cr_9,\nonumber\\
 w_{42}=h_1 ,\hspace{2mm}w_{73}=h_2,  \nonumber\\
 w_{24}=h_3 ,\hspace{2mm}w_{86}=h_4, \nonumber\\
 w_{68}=h_5 ,\hspace{2mm}w_{37}=h_6. \nonumber\\
\end{eqnarray}

In the quantum Hamiltonian formalism, the bulk transition matrix is then given by
\bel{2-16b}
h_{k,k+1}= - \left( \ba{ccccccccc}
. & 0 & 0 & 0 & 0 & 0 & 0 & 0 & 0\\0 & . & r_3 & h_3 & 0 & 0 & r_4 & 0 & 0 \\ 
0 & cr_1 &. & cr_4 &0 &0 &h_6 &0 &0 \\0 & h_1 & r_2 &. & 0 & 0 & r_1 & 0 & 
0\\0 & 0 &0 & 0&. & r_6 & 0 & r_5 & r_7\\0& 0 & 0 & 0 & cr_5  &. & 0 & h_5& 
r_9\\0 & cr_2& h_2& cr_3& 0 &0 &.& 0& 0\\ 0& 0& 0& 0& cr_6& h_4 & 0 & . & 
r_8\\
0 &0 & 0  & 0 & c^2r_7 & cr_8 & 0 & cr_9 & . \\ \ea
\right)_{k,k+1}.
\ee

\subsection{Product measure}

With (\ref{marginal}) the homogeneous product measure has the form
\begin{equation}\label{steadystate}
\ket{P^*}=\frac{1}{{\nu}^L}\left( \ba{c} 1 \\ z\\cz \ea \right)^{\otimes L}.
\end{equation}
It is convenient to define
\bel{hhat}
\hat{h}_{i,i+1} = h_i - [E(\hat n^A_{i}-\hat n ^A_{i+1}) +E'(\hat n^B_{i} -\hat 
n^B_{i+1})]
\ee
where $E,E'$ are arbitrary constants and 
 $\hat n_A$ and $\hat n_B$ are number 
operators with eigenvalue 1 if a particle of the respective species
is present and 0 otherwise. Furthermore we define modified boundary matrices
\bel{bhat}
\hat{b}_{1} = b_1 + E \hat n^A_{1}+E'\hat n^B_{1}, \quad 
\hat{b}_{L} = b_L - E \hat n^A_{L}-E'\hat n^B_{L}.
\ee
This allows us to rewrite the quantum Hamiltonian as
\bel{H}
H = \hat{b}_{1} + \sum_{i=1}^{L-1} \hat{h}_{i,i+1} + \hat{b}_{L}.
\ee
The eigenvalue equation (\ref{eq:r1}) may be rewritten
\begin{equation}\label{eq:r1a}
0 =\hat{h}_{i,i+1}\ket{P^*} = (\hat{b}_{1} + g)\ket{P^*} = (\hat{b}_{L}-g)\ket{P^*}.
\end{equation}
with a further arbitrary constant $g$.

This trick allows us to determined the conditions on the rates that ensure
that (\ref{steadystate}) actually is stationary. For the bulk rates 
(\ref{eq:r1a}) yields
\bea
&E&=h_3-h_1+c(r_3+r_4-r_1-r_2),\nn
&E'&=h_6-h_2+r_1+r_4-r_2-r_3.\label{eq:r31}
\eea
Furthermore, some algebra shows that the bulk rates must satisfy the 
following condition for stationarity 
\begin{equation}
h_6-h_2+h_1-h_3+h_4-h_5+(1+c)(r_1-r_3)+(1-c)(r_4-r_2)+c(r_8-r_9)+r_6-r_5=0.\label{eq:stationarity}
\end{equation}
In order to satisfy the eigenvalue equation at the boundaries the terms
involving $E,E'$ must cancel. For the
left boundary this yields the two relations 
\begin{eqnarray}
[h_1-h_3+c(2(r_2-r_4)+h_2-h_6)]z&=&\gamma_1 z\nu+\gamma_2 
cz\nu-(\alpha_1+\alpha_2)\nu\nonumber\\
&=&-\beta_1 z\nu-\beta_2 cz\nu+(\delta_1+\delta_2)\nu, \nonumber \\
\end{eqnarray}
and similarly at the right boundary
\begin{eqnarray}
[(-r_5+r_6+c(r_8-r_9)+h_4-h_5)z-r_1+r_2+r_3-r_4+h_2-h_6]cz\nonumber\\
=(\gamma_2+\gamma_3)cz\nu-\alpha_2\nu-\alpha_3z(1+z)\nu\nonumber\\
=-(\beta_2+\beta_3)cz\nu+\delta_2\nu+\delta_3z(1+z)\nu.
\end{eqnarray}
These relations define a model for which the product measure with constant
fugacity $z$ is stationary. The fugacity is determined by its boundary value
encoded in the boundary rates.

\subsection{Fugacity gradient}

Now we generalize the model to allow for different fugacities $z_1,z_2$
at the two boundaries. The product measure is then no longer stationary
and there is no general principle that would constrain the form
of the stationary distribution. However, in principle its properties can be 
calculated from the studying the time evolution of the system starting
from some initial distribution.

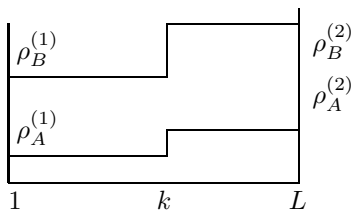
\begin{figure}[h]
\begin{center}
\begin{picture}(120,70)
\put(10,10){\line(0,1){60}}
\put(10,10){\line(1,0){110}}
\put(10,20){\line(1,0){60}}
\put(70,20){\line(0,1){10}}
\put(70,30){\line(1,0){50}}
\put(120,10){\line(0,1){66}}
\put(10,50){\line(1,0){60}}
\put(70,50){\line(0,1){20}}
\put(70,70){\line(1,0){50}}
\put(10,0){\footnotesize 1}
\put(66,0){\footnotesize $k$}
\put(116,0){\footnotesize $L$}
\put(125,40){\footnotesize $\rho_A^{(2)}$}
\put(13,27){\footnotesize $\rho_A^{(1)}$}
\put(125,60){\footnotesize $\rho_B^{(2)}$}
\put(13,57){\footnotesize $\rho_B^{(1)}$}
\end{picture}
\caption[profig]{Coarse grained density profiles of a shock measure with shock 
between 
sites $k,k+1$.}
\label{profig}
\end{center}
\end{figure}

In general, solving for the dynamics of a many-particle system is a 
much harder task than determining its stationary distribution. However,
guided by previous experience \cite{Kreb03} we make as ansatz an
initial distribution which is a shock measure connecting the two
boundary fugacities. The representation of the shock measure here is
\begin{equation}
\ket k=\frac{1}{\nu_1^k\nu_2^{(L-k)}}\left( \ba{c} 1 \\ z_1 \\ cz_1 \ea 
\right)^{\otimes k}\otimes \left( \ba{c} 1 \\ z_2 \\ cz_2 \ea 
\right)^{\otimes L-k}.  
\end{equation}          
On a coarse-grained scale the density profile corresponding to this measure
has a jump discontinuity, see Fig. \ref{profig}. We search for
conditions on the rates such that
\begin{equation}
\frac {d}{dt} \ket k = d_1 \ket{k-1} +d_2 \ket{k+1}-(d_1+d_2)\ket k.
\label{eq:r4}
\end{equation}
This implies that the family of shock measures labelled by the
shock position $k$ is closed under the time evolution of the many-particle
system. Physically this behaviour corresponds to a random walk
of the shock with hopping rates $d_1$, ($d_2$) to the left (right).

In order to have the random walk equation (\ref{eq:r4}) for the shock, 
one replaces the left hand side by the (negative) quantum Hamiltonian
in the form (\ref{H}). Then
in each branch of the shock one has $\hat{h}_{i,i+1} \ket k = 0$,
except for $i=k$. Stationarity at the boundaries implies 
\be
b_1\ket{P^*}=(-E\hat n^A_1-E'\hat n^B_1+g_1)\ket{P^*},
\end{equation}
\begin{equation}
b_L\ket{P^*}=(E\hat n^A_L+E'\hat n^B_L-g_2)\ket{P^*},
\end{equation}
where $g_1$ and $g_2$ are obtained using (\ref{eq:r31}) as
\bea
g_1&=&E\frac{z_1}{\nu_1}+E'\frac{cz_1}{\nu_1}=(1+c)(p-q)\frac{z_1}{\nu_1}\nn
&=&\alpha_1+\alpha_2-(\gamma_1+c\gamma_2)z_1,
\eea
\bea
g_2&=&E\frac{z_2}{\nu_2}+E'\frac{cz_2}{\nu_2}=(1+c)(p-q)\frac{z_2}{\nu_2}\nn
&=&-(\delta_1+\delta_2)+(\beta_1+c\beta_2)z_2.
\eea

The random walk condition for the shock thus leads to 9 
equations  
\begin{equation}
(-\hat{h_k}+d_1+d_2-g_1+g_2)\ket k-d_1\ket{k-1} - d_2\ket{k+1}=0
\label{randomwalkcondition}
\end{equation}
for the bulk rates. Three of these conditions are 
fulfilled without any constraint on the rates, leaving the following 6 equations :
\bea
& &S-d_1\frac{\nu_1}{\nu_2}-d_2\frac{\nu_2}{\nu_1}=0,\\
& &(cr_4+h_3)(z_1-z_2)+Sz_2-d_1z_2\frac{\nu_1}{\nu_2}-d_2z_1\frac{\nu_2}
{\nu_1}=0,\\
& &(r_4+h_6)(z_1-z_2)+Sz_2-d_1z_2\frac{\nu_1}{\nu_2}-d_2z_1\frac{\nu_2}
{\nu_1}=0,\\
& &(cr_2+h_1)(z_2-z_1)+Sz_1-d_1z_2\frac{\nu_1}{\nu_2}-d_2z_1\frac{\nu_2}
{\nu_1}=0,\\
& &S-d_1\frac{z_2\nu_1}{z_1\nu_2}-d_2\frac{z_1\nu_2}{z_2\nu_1}=0,\\
& &(r_2+h_2)(z_2-z_1)+Sz_1-d_1z_2\frac{\nu_1}{\nu_2}-d_2z_1\frac{\nu_2}
{\nu_1}=0,
\eea
where
\begin{equation}
S=d_1+d_2+g_2-g_1.
\end{equation}
Solving the above equations leads to three independent relations between bulk rates 
and densities
\begin{equation}
h_3+cr_4=h_6+r_4\equiv p,\label{eq:rs1}
\end{equation}
\begin{equation}
h_1+cr_2=h_2+r_2\equiv q,\label{eq:rs2}
\end{equation}
\begin{equation}
\frac{p}{q}=\frac{z_2}{z_1},
\end{equation}
and two relations
\begin{equation}
d_1=q\frac{\nu_2}{\nu_1},
\end{equation}
\begin{equation}
d_2=p\frac{\nu_1}{\nu_2},
\end{equation}
that express the shock hopping rates in terms
of the hopping rates of the model and the fugacities of the shock.
On this parameter manifold the stationarity condition 
(\ref{eq:stationarity}) reduces to 
\begin{equation}
h_4-h_5+(1+c)(r_1-r_3)+c(r_8-r_9)+r_6-r_5=0.
\end{equation}

The shock performs a random walk for a specific ratio
of the boundary fugacities, or, equivalently, at some
specific strength of the driving force encoded in the
particle hopping rates. Thus shock mean velosity $v_s$ in terms of vacancy density and hopping rates is
\begin{equation}
v_s=\frac{q\nu_{2}^{2}-p\nu_{1}^2}{\nu_{1}\nu_{2}},
\end{equation}
and its diffusion coefficient as long as the shock is far from the boundaries is
\begin{equation}
D_s=\frac{p\nu_{1}^2+q\nu_{2}^2}{2\nu_{1}\nu_{2}}.
\end{equation}


From the shock hopping rates and its biased random walk dynamics
we can read off the stationary distribution of the system for 
different boundary densities. This is a linear combination of shock 
measures
\be
\ket{P^\ast} \propto \sum_k \left(\frac{d_1}{d_2}\right)^k \ket{k}.
\ee
For $d_1>d_2$ (bias to the right) the stationary shock position is in the
vicinity of the right boundary, leaving the system in a phase of low
density. Conversely, for $d_1<d_2$, the system is in a high-density
phase. At $d_1=d_2$ the system undergoes a first-order nonequilibrium
transition \cite{Kolo98}. Here the shock has no bias and can be found
with equal probability anywhere on the lattice. The stationary density
profile is linear, but a typical particle configuration has two different
regions of constant (but fluctuating) density. The density jumps
quickly from one density to another in some small region of the lattice.

\subsection{Steady state current}

In order to make contact with the ASEP we calculate the stationary 
current for this model. In order to identify the current we first calculate the 
equation of motion for the expected local particle densities, 
\bea
\frac{d}{dt}\langle n_k^A \rangle
&=&-(h_1+cr_1+cr_2)\langle n_{k-1}^0 n_k^A \rangle+h_3\langle n_{k-1}^A n_k^0 
\rangle+(h_4-r_6)\langle n_{k-1}^A n_k^B \rangle \nn
&&-(h_5+cr_9)\langle n_{k-1}^B n_k^A \rangle +r_3\langle n_{k-1}^0 n_k^B 
\rangle+ r_4\langle n_{k-1}^B n_k^0 \rangle \nn
&&-c(r_5+cr_7)\langle n_{k-1}^A n_k^A \rangle+(r_7+r_8)\langle n_{k-1}^B n_k^B 
\rangle+h_1\langle n_{k}^0 n_{k+1}^A \rangle \nn
&&-(h_3+cr_3+cr_4)\langle n_{k}^A n_{k+1}^0 \rangle-(h_4+cr_8)\langle n_{k}^A 
n_{k+1}^B \rangle \nn
&&+(h_5+r_5)\langle n_{k}^B n_{k+1}^A \rangle+r_1\langle n_{k}^B n_{k+1}^0 
\rangle-r_2\langle n_{k}^0 n_{k+1}^B\rangle\nn
&&-c(r_6+cr_7)\langle n_{k}^A n_{k+1}^A \rangle+(r_7+r_9)\langle n_{k}^B 
n_{k+1}^B \rangle,\nn
\eea
\bea
\frac{d}{dt}\langle n_k^B \rangle&=&-(h_2+r_2+r_3)\langle n_{k-1}^0 n_k^B 
\rangle+h_6\langle n_{k-1}^B n_k^0 \rangle-(h_4+r_6)\langle n_{k-1}^A n_k^B 
\rangle\nn
&&+(h_5+cr_9)\langle n_{k-1}^B n_k^A \rangle +cr_1\langle n_{k-1}^0 n_k^A 
\rangle+cr_4\langle n_{k-1}^A n_k^0 \rangle \nn
&&+c(r_5+cr_7)\langle n_{k-1}^A n_k^A \rangle-(r_7+r_8)\langle n_{k-1}^B n_k^B 
\rangle +h_2\langle n_{k}^0 n_{k+1}^B \rangle \nn
&&-(h_6+r_1+r_4)\langle n_{k}^B n_{k+1}^0 \rangle+(h_4+cr_8)\langle n_{k}^A 
n_{k+1}^B \rangle\nn
&&-(h_5+r_5)\langle n_{k}^B n_{k+1}^A \rangle+cr_2\langle n_{k}^0 
n_{k+1}^A\rangle+cr_3\langle n_{k}^A n_{k+1}^0 \rangle\nn
&&+c(r_6+cr_7)\langle n_{k}^A n_{k+1}^A \rangle-(r_7+r_9)\langle n_{k}^B 
n_{k+1}^B \rangle.\nn
\eea

This can be written in terms of $A$ and $B$ particle current
\be
\frac{d}{dt}\langle n_k^A\rangle=j^A_{k-1}-j_k^A+S_k,
\ee
\be
\frac{d}{dt}\langle n_k^B\rangle=j^B_{k-1}-j_k^B-S_k,
\ee
where the source term
\bea
S_k&=&(cr_1-\frac{cr_2}{2})\langle n_{k-1}^0n_k^A\rangle+(\frac{r_2}
{2}+r_3)\langle n_{k-1}^0n_k^B\rangle+\frac{r_4}{2}\langle 
n_{k-1}^Bn_k^0\rangle\nn
&&-\frac{cr_4}{2}\langle n_{k-1}^An_k^0\rangle-(cr_5+c^2r_7)\langle 
n_{k-1}^An_k^A\rangle+r_6\langle n_{k-1}^An_k^B\rangle\nn
&&+(r_7+r_8)\langle n_{k-1}^Bn_k^B\rangle-cr_9\langle 
n_{k-1}^Bn_k^A\rangle+(r_1+\frac{r_4}{2})\langle n_k^Bn_{k+1}^0\rangle\nn
&&+\frac{r_2}{2}\langle n_k^0n_{k+1}^B\rangle-\frac{cr_2}{2}\langle 
n_k^0n_{k+1}^A\rangle-(cr_3+\frac{cr_4}{2}) \langle n_k^An_{k+1}^0\rangle\nn
&&+r_5\langle n_k^Bn_{k+1}^A\rangle-(cr_6+c^2r_7)\langle 
n_k^An_{k+1}^A\rangle+(r_7+r_9)\langle n_k^Bn_{k+1}^B\rangle \nn
&&-cr_8\langle n_k^An_{k+1}^B\rangle.\nn
\eea
expresses the fact that the individual particle densities are not
conserved. The particle currents are given  by the expectations
\bea
j^A_k&=&-(h_1+\frac{cr_2}{2})\langle n_k^0n_{k+1}^A\rangle+(h_3+\frac{cr_4}
{2})\langle n_k^An_{k+1}^0\rangle+h_4\langle n_k^An_{k+1}^B\rangle\nn
&&-h_5\langle n_k^Bn_{k+1}^A\rangle-\frac{r_2}{2}\langle 
n_k^0n_{k+1}^B\rangle+\frac{r_4}{2}\langle n_k^Bn_{k+1}^0\rangle,\nn
\eea
\bea
j^B_k&=&-(h_2+\frac{r_2}{2})\langle n_k^0n_{k+1}^B\rangle+(h_6+\frac{r_4}
{2})\langle n_k^Bn_{k+1}^0\rangle-h_4\langle n_k^An_{k+1}^B\rangle\nn
&&+h_5\langle n_k^Bn_{k+1}^A\rangle-\frac{cr_2}{2}\langle 
n_k^0n_{k+1}^A\rangle+\frac{cr_4}{2}\langle n_k^An_{k+1}^0\rangle.\nn
\eea

By adding the two individual currents we find the total 
particle current to be given by
\bea
j_k&=&j_k^A+j_k^B \nn
&=&-h_1\langle n_k^0n_{k+1}^A\rangle+h_3\langle 
n_k^An_{k+1}^0\rangle-h_2\langle n_k^0n_{k+1}^B\rangle+h_6\langle 
n_k^Bn_{k+1}^0\rangle\nn
&&-r_2\langle n_k^0n_{k+1}^B\rangle-cr_2 \langle 
n_k^0n_{k+1}^A\rangle+cr_4\langle n_k^An_{k+1}^0\rangle +r_4\langle 
n_k^Bn_{k+1}^0\rangle.\nn
\eea
In the steady state we obtain  
\begin{equation}
j^*=\frac{h_3-h_1+c(h_6-h_2)+2c(r_4-r_2)}{1+c} \rho(1-\rho),
\end{equation}
where $\rho$ is the average density (\ref{density}).
This can be written  in terms of $E$ and $E'$ 
\begin{equation}
j^*=\frac{E+cE'}{1+c}\rho(1-\rho).
\end{equation}
This is the well-known parabolic current-density relation of 
the ASEP \cite{Ligg99,Schu00} where the density-independent prefactor
plays the role of the hopping bias. In fact, on the special manifold
which gives rise to the random walk of the shock we find, using 
(\ref{eq:rs1})-(\ref{eq:rs2}), the simpler expression
\begin{equation}
j^*=(p-q)\rho(1-\rho).
\end{equation}

\section{Conclusions}

We have found that three-states lattice gases with a single local
conservation law can be classified into two families, one where the
conserved quantity is a linear function of the occupation variable,
another where the function is degenerate, i.e., takes the same
value for two different states. Nonlinear nondegenerate functions
lead to two independently conserved quantities.

The degenerate linear conservation describes a class of asymmetric 
exclusion processes with a binary internal degree of freedom. 
We have identified constraints on the transition rates such that
the stationary distribution is a product measure, parametrized
by the nonequilibrium analog of the fugacity. For open systems with
different boundary fugacities we have found a complete list of models
where the shock performs a biased random walk on the lattice. For these 
systems we have detailed knowledge about the microscopic structure of the 
shock. As in other models studied previously (see \cite{Taba06b} and references
therein) these shocks are intrinsically maximally sharp and behave
like collective single-particle excitations already on the lattice scale --
not only after coarse-graining where all the microscopic features of the
shock are lost. Apparently this enormous reduction in the number of
dynamical degrees of freedom in a subspace of the stochastic
dynamics appears more frequently than previously
suggested \cite{Bala01}. 

An immediate consequence of the random walk dynamics of the shock
is the existence of a first order boundary-induced phase transition
which occurs if the boundary fugacities reverse the mean shock velocity.
Away from this special manifold
our result for the sharpness of the shock suggest that
finite systems with lattice size of the
order $10$ can be well described by the domain wall theory for first order 
boundary-induced phase transitions, \cite{Kolo98,Popk99}, with
limitations analogous to those obtained from the exact results of Ref.
\cite{Gier05}. 

It is intriguing
that the maximal sharpness appears at some specific value of the
driving force or, equivalently, ratio of boundary fugacities. It
would be interesting to investigate
whether such a field-induced
sharpening of the interface is a special property of lattice models
or can appear also in continuum systems such as the recently studied
mass transfer models \cite{Evan04,Evan06}. It is also an open problem
whether there can be an analogous reduction of the shock dynamics
to a random walk problem
in exclusion processes where the stationary distribution does not
factorize \cite{Katz84,Anta00}.

\subsection*{Acknowledgments}
FT would like to thank R.J.Harris for useful discussions.

\newpage

\bibliographystyle{unsrt}

\end{document}